\documentclass[aps,prd,preprintnumbers,superscriptaddress,nofootinbib,notitlepage,floatfix,10pt,twocolumn]{revtex4-2}
\usepackage[pdftex]{graphicx}
\usepackage{bm,latexsym,amsmath,amssymb,amsfonts,mathrsfs}
\usepackage{color}
\allowdisplaybreaks[1]
\usepackage[pdftex,colorlinks=true,linkcolor=blue,citecolor=cyan,backref=page]{hyperref}
\newcommand*{\D}{\mathrm{d}}

\begin{document}
\title{Gravitomagnetic tidal response of relativistic stars in partially screened scalar-tensor theories}

%
\author{Tsutomu~Kobayashi}
\email[Email: ]{tsutomu@rikkyo.ac.jp}
\affiliation{Department of Physics, Rikkyo University, Toshima, Tokyo 171-8501, Japan}
%
\begin{abstract}
In scalar-tensor theories beyond Horndeski, the Vainshtein screening mechanism is only partially effective inside astrophysical bodies.
We investigate the potential to detect this partial breaking of Vainshtein screening through the tidal response of fluid bodies.
We calculate the gravitomagnetic tidal Love numbers
in a specific model of degenerate higher-order scalar-tensor gravity and analyze how deviations from general relativity depend on parameters governing the breaking of Vainshtein screening in the weak-gravity regime.
For fixed parameter values, the relative deviations increase with higher multipoles
and larger compactness.
However, we demonstrate that these parameters alone are insufficient to fully characterize the tidal
response of relativistic bodies in scalar-tensor theories beyond Horndeski.
\end{abstract}
\preprint{RUP-25-1}
\maketitle

\section{Introduction}\label{sec:intro}

The response of a self-gravitating body to an applied external field
depends on its internal structure and the gravitational theory,
which is characterized by a set of tidal Love numbers~\cite{Will:2016sgx}.
Love numbers were first introduced in Newtonian gravity~\cite{Love:1909cid}.
The extension of Love numbers
to general relativity (GR) was initiated by~\cite{Flanagan:2007ix,Hinderer:2007mb}
and their notion was made more precise in~\cite{Damour:2009vw,Binnington:2009bb}.
For nonrotating bodies, the relativistic Love numbers are classified into two types
based on parity: gravitoelectric Love numbers associated with an even-parity tidal field 
and gravitomagnetic Love numbers associated with an odd-parity tidal field,
with the latter emerging as a purely relativistic effect.
The relativistic tidal Love numbers lay the foundation for 
probing the equation of state of nuclear matter in neutron stars
with gravitational waves from compact binaries~\cite{Hinderer:2009ca,LIGOScientific:2018cki}.

The tidal response is also sensitive to the underlying gravitational theory, potentially enabling us to
test GR through the tidal deformability of neutron stars.
Among various possibilities, scalar-tensor theories have the simplest field content
consisting of one scalar and two tensor degrees of freedom and offer us
a theoretically consistent and physically well-motivated framework to explore deviations from GR.
Of particular interest are theories that incorporate a mechanism to screen
the effects of the scalar degree of freedom near a matter source,
thereby evading existing tests of gravity within the solar system.
The tidal response of neutron stars in scalar-tensor theories has been discussed in Refs.~\cite{Pani:2014jra,Yazadjiev:2018xxk,Saffer:2021gak,Brown:2022kbw,Creci:2023cfx,Diedrichs:2025vhv}.
However, to the best of our knowledge,
no analysis in that direction has been carried out so far for scalar-tensor theories equipped with
the Vainshtein screening mechanism.
(See Ref.~\cite{Babichev:2013usa} for a review of the Vainshtein mechanism.)
Scalar-tensor theories in the Horndeski family~\cite{Horndeski:1974wa,Deffayet:2011gz,Kobayashi:2011nu},
though not all,
typically feature this mechanism due to nonlinear derivative interactions of the scalar
field~\cite{Kimura:2011dc,Narikawa:2013pjr,Koyama:2013paa}.\footnote{The very recent work
on the tidal deformability of neutron stars in the Horndeski theory~\cite{Diedrichs:2025vhv} has
considered concrete models without the Vainshtein mechanism.}
This screening mechanism operates so efficiently in the Horndeski family that
practically no deviations from GR are expected
in the Vainshtein regime, i.e. inside the Vainshtein radius.
However, in scalar-tensor theories beyond Horndeski~\cite{Zumalacarregui:2013pma,Gleyzes:2014dya},
Vainshtein screening is only partially effective inside a region filled with matter,
although it is complete outside~\cite{Kobayashi:2014ida},
leading to astrophysical tests of gravity within this class
through the modified internal structure of astrophysical bodies~\cite{Koyama:2015oma,Saito:2015fza}.
This motivates us to study the tidal deformability of relativistic stars in partially screened
scalar-tensor theories, given that the tidal response is sensitive to both the internal structure of objects and
the underlying theory of gravity.
See Ref.~\cite{Banerjee:2020rrd} for related work exploring a somewhat similar direction in the Newtonian limit.

Scalar-tensor theories beyond Horndeski have been systematically constructed and classified
under the name of degenerate higher-order scalar-tensor (DHOST)
theories~\cite{Langlois:2015cwa,Crisostomi:2016czh,BenAchour:2016fzp}.
See~\cite{Langlois:2018dxi,Kobayashi:2019hrl} for reviews.
The breaking of Vainshtein screening generically occurs in
DHOST theories~\cite{Crisostomi:2017lbg,Langlois:2017dyl,Dima:2017pwp}.
The Tolman-Oppenheimer-Volkoff (TOV) system for the relativistic stellar structure in DHOST theories
has been considered in Refs.~\cite{Babichev:2016jom,Sakstein:2016oel,Chagoya:2018lmv,Kobayashi:2018xvr}.
In this paper, we study the gravitomagnetic tidal response of relativistic stars
in the Vainshtein regime of DHOST theories
and calculate the associated tidal Love numbers, illustrating how they depend on the parameters of
the underlying theory of modified gravity governing the partial breaking of Vainshtein screening.
Enriching our understanding of the tidal response of relativistic stars in modified gravity would help us develop accurate waveform models and break the degeneracy between uncertainties in the nuclear equation of state and gravity beyond GR.

The paper is organized as follows.
In the next section, we provide the basic equations that determine
our unperturbed configuration, following closely the previous work~\cite{Kobayashi:2018xvr}.
We also see how the Newtonian limit can differ from the standard behavior inside a fluid body.
Section~\ref{sec:Odd-pert} details the derivation of the equation for the odd-parity perturbations
in the static limit
and the basic procedure to compute the gravitomagnetic Love numbers in DHOST theories. 
In Sec.~\ref{sec:model}, we give a family of concrete DHOST models that we use in calculating 
the Love numbers.
Some discussions on the unperturbed configurations are also provided.
Our main results are presented in Sec.~\ref{sec:Love}, clarifying how the gravitomagnetic
Love numbers depend on the modified gravity parameters.
Finally, we draw our conclusions in Sec~\ref{sec:conclusions}.

\section{TOV system in DHOST theories}\label{sec:TOV}

\subsection{Quadratic DHOST theories}

The action of quadratic DHOST theories is given by~\cite{Langlois:2015cwa}
\begin{align}
        S&=\int\D^4x\sqrt{-g}\left[
                f(X)\mathcal{R}+\sum_{I=1}^5 A_I(X)L_I
        \right]+S_{\textrm{m}},\label{eq:action-DHOST}
\end{align}
where $\mathcal{R}$ is the Ricci scalar, $X:=-\phi^\mu\phi_\mu/2$,
\begin{align}
        &L_1:=\phi_{\mu\nu}\phi^{\mu\nu},
        \quad L_2:=(\Box\phi)^2,
        \quad L_3:=\Box\phi \phi^\mu\phi_{\mu\nu}\phi^\nu,
        \notag \\ &
        L_4:=\phi^\mu\phi_{\mu\rho}\phi^{\rho\nu}\phi_\nu,
        \quad 
        L_5:=\left(\phi^\mu\phi_{\mu\nu}\phi^\nu\right)^2,
\end{align}
with the notations $\phi_\mu:=\nabla_\mu\phi$ and $\phi_{\mu\nu}:=\nabla_\mu\nabla_\nu\phi$,
and $S_{\textrm{m}}$ is the action for a perfect fluid minimally coupled to gravity.
The functions $f$ and $A_I$ must obey the degeneracy conditions so that the system is
described by two tensor and one scalar dynamical degrees of freedom.
We assume the shift symmetry of $\phi$,
and hence $f$ and $A_I$ are dependent only on $X$.
One may add the terms of the form $F_0(X)+F_1(X)\Box\phi$ to the Lagrangian
without changing the number of dynamical degrees of freedom.
However, in this paper, we only study the Vainshtein regime of quadratic DHOST theories where
these terms can be ignored~\cite{Crisostomi:2017lbg,Langlois:2017dyl,Dima:2017pwp}.

In light of the joint observation of GW170817 and GRB 170817A~\cite{LIGOScientific:2017vwq,LIGOScientific:2017zic,LIGOScientific:2017ync},
we focus on the subset of DHOST theories in which the speed of gravitational waves is equal to that of light.
This subset satisfies $A_1=0$~\cite{deRham:2016wji}.
The degeneracy conditions are then given by~\cite{Langlois:2015cwa}
\begin{align}
        A_2&=-A_1=0,
        \\ 
        A_4&=-\frac{1}{2f}\left(
                2fA_3-3f_X^2-2Xf_XA_3+X^2A_3^2
        \right),
        \\ 
        A_5&=-\frac{A_3}{f}\left(f_X+XA_3\right),
\end{align}
while $f$ and $A_3$ are free.
Here and hereafter, $f_X$ denotes $\partial f/\partial X$.

The field equations derived from the action~\eqref{eq:action-DHOST} are of the form
\begin{align}
        \frac{2}{\sqrt{-g}}\frac{\delta S}{\delta g^{\mu\nu}}
        =\mathcal{E}_{\mu\nu}-T_{\mu\nu}&=0,\label{eom:metric}
        \\ 
        \nabla_\mu \mathcal{J}^\mu&=0,\label{eom:scalar}
\end{align}
where 
$T_{\mu\nu}:=-(2/\sqrt{-g})\delta S_{\textrm{m}}/\delta g^{\mu\nu}$
is the energy-momentum tensor for the perfect fluid
and 
$\mathcal{J}^\mu:=(1/\sqrt{-g})\delta S/\delta\phi_\mu$
is the shift current.
The energy-momentum tensor satisfies the conservation equations (the hydrodynamical equations)
\begin{align}
        \nabla_\nu T_\mu^\nu=0.
        \label{eq:conservation-eq-fluid}
\end{align}
The hydrodynamical equations are identical to those in GR because
the fluid is assumed to be minimally coupled with gravity.

\subsection{Governing equations for the TOV system}\label{subsec:TOV-system}

Let us consider a spherically symmetric solution in DHOST theories,
which will be used as an unperturbed background configuration.
The metric and the scalar field are taken to be 
\begin{align}
        \bar g_{\mu\nu}\D x^\mu \D x^\nu 
        &=-e^{\nu(r)}\D t^2+e^{\lambda(r)}\D r^2+r^2\D\Omega^2,
        \\ 
        \phi&=t+\psi(r),
\end{align}
with $\D\Omega^2=\D\theta^2+\sin^2\theta\D\varphi^2$.
Although the metric is assumed to be static, $\phi$ can depend linearly on $t$ thanks to the shift symmetry.
Notice that $\phi$ has the dimension of time in our convention.
The energy-momentum tensor is given by
\begin{align}
        \bar T_\mu^\nu =\left(\rho+p\right)\bar u_\mu \bar u^\nu +p\delta_\mu^\nu,
\end{align}
with the four-velocity
\begin{align}
        \bar u^\mu = \left(e^{-\nu/2},0,0,0\right),
\end{align}
where $\rho=\rho(r)$ and $p=p(r)$ are the energy density and the pressure, respectively.
They are related through the equation of state: $p=p(\rho)$.

Although the field equations appear to be of higher order in DHOST theories,
the degeneracy of the system allows one to reduce
the number of derivatives by combining different components of the field equations.
In the present case, the procedure was elaborated in Ref.~\cite{Kobayashi:2018xvr},
which is reviewed in Appendix~\ref{app:derivation-TOV} for completeness.
One arrives in the end at the following set of equations:
\begin{align}
        e^{\lambda}&=\mathcal{F}_\lambda(\nu,\nu',X,X',p),
        \label{eq:elambda}
        \\ 
        X'&=\mathcal{F}_1(\nu,X,\rho,p)\nu'+\frac{\mathcal{F}_2(\nu,X,\rho,p)}{r},
        \label{eq:dX}
        \\ 
        \nu'&=\mathcal{F}_3(\nu,X,\rho,\rho',p),
        \label{eq:dnu}
\end{align}
supplemented with the radial component of the hydrodynamical equations~\eqref{eq:conservation-eq-fluid},
\begin{align}
        p'=-\frac{\nu'}{2}(\rho+p),\label{eq:hydro-r}
\end{align}
where a prime denotes differentiation with respect to $r$.
Here, it is more convenient to use $X=[e^{-\nu}-e^{-\lambda}(\psi')^2]/2$
rather than $\psi'$.
Although the explicit expressions for $\mathcal{F}_\lambda$, $\mathcal{F}_1$, $\mathcal{F}_2$, and $\mathcal{F}_3$
are extremely messy, it is straightforward to reproduce them by following the steps outlined in Appendix~\ref{app:derivation-TOV}
with the help of \textit{Mathematica}.
See Ref.~\cite{Kobayashi:2018xvr}
and Appendix~\ref{app:derivation-TOV}
for the explicit expressions,
but notice that our definition of $X$ differs by a factor of $-1/2$ from that in Ref.~\cite{Kobayashi:2018xvr}.
Given the equation of state $p=p(\rho)$, we can remove $p$ (or $\rho$) from the above equations.

\subsection{External solution, boundary conditions at the center, and matching at the surface}

Setting $p=\rho=0$ in the external region, Eqs.~\eqref{eq:elambda}--\eqref{eq:dnu} are simplified to
\begin{align}
        e^\lambda=1+r\nu',\quad X'=0,\quad \nu'=-\frac{1}{r}+\frac{e^{-\nu}}{2rX}.
\end{align}
By requiring that $\psi'(r)\to 0$ and $\nu(r)\to 0$ as $r\to \infty$,
we find~\cite{Kobayashi:2018xvr}
\begin{align}
        e^\nu&=e^{-\lambda}=1-\frac{2\mu}{r},
        \label{eq:ext-soln-met}
        \\ 
        X&=\frac{1}{2}.\label{eq:ext-soln-X}
\end{align}
Here, $\mu$ is an integration constant that is determined by matching
the external and internal solutions at the stellar surface $r=R$.
The external solution is thus given by the Schwarzschild solution,
which manifests complete Vainshtein screening in the external region.

To integrate the equations in the internal region, let us provide the boundary conditions imposed at the center.
In the vicinity of the center, we have a series expansion of the form, 
\begin{align}
        \nu&=\nu_c+\frac{\nu_2}{2}r^2+\dots,\label{bc:1}
        \\ 
        X&=\frac{e^{-\nu_c}}{2}\left(1+\frac{X_2}{2}r^2+\dots\right),\label{bc:2}
        \\ 
        \rho&=\rho_c+\frac{\rho_2}{2}r^2+\dots,\label{bc:3}
        \\ 
        p&=p_c+\frac{p_2}{2}r^2+\dots,\label{bc:3p}
\end{align}
where $\psi'(0)=0$ is assumed.
Expanding Eq.~\eqref{eq:elambda} around $r=0$, we see that
\begin{align}
        \lambda=\frac{\lambda_2}{2}r^2+\dots,\label{bc:4}
\end{align}
where $\lambda_2$ can be expressed in terms of $\rho_c$, $p_c$, $\nu_c$, $\nu_2$, and $X_2$.
Expanding then Eqs.~\eqref{eq:dX},~\eqref{eq:dnu}, and~\eqref{eq:hydro-r} around $r=0$,
one finds the expressions for $\nu_2$, $X_2$, and $p_2$ in terms of $\rho_c$, $p_c$, and $\nu_c$.
The results are summarized as follows:
\begin{align}
        \nu_2&=8\pi G_c \Biggl[
                \frac{\rho_c+3p_c}{3}-\frac{(\tilde{\alpha}_H+\tilde\beta_1)^2}{\tilde\alpha_H+2\tilde\beta_1}\rho_c
                \notag \\ & \quad 
                -\frac{(2\tilde\alpha_H+\tilde{\beta}_1)(\tilde{\alpha}_H+3\tilde{\beta}_1)}{\tilde\alpha_H+2\tilde\beta_1}p_c
                \Biggr],
                \\
        X_2&=-\frac{8\pi G_c}{\tilde\alpha_H+2\tilde\beta_1}\left[
                (2\tilde\alpha_H+3\tilde\beta_1)\rho_c 
                +3(\tilde\alpha_H+3\tilde\beta_1)p_c 
        \right],
        \\ 
        \lambda_2&=-\frac{p_c}{f(e^{-\nu_c}/2)}+2\nu_2-2\tilde\alpha_H X_2,
        \\ 
        p_2&=-\frac{\nu_2}{2}(\rho_c+p_c),
\end{align}
with 
\begin{align}
        \tilde\alpha_H&:=\left.-\frac{2Xf_X}{f}\right|_{X=e^{-\nu_c}/2},
        \\ 
        \tilde{\beta}_1&:=\left.\frac{X}{f}\left(f_X+XA_3\right)\right|_{X=e^{-\nu_c}/2},
        \\ 
        8\pi G_c&:=\frac{1}{2f(e^{-\nu_c}/2)(1-\tilde{\alpha}_H-3\tilde{\beta}_1)}.
\end{align}
It is interesting to compare these expressions with Eqs.~\eqref{eq:def-eft} and~\eqref{def:GNewton}.

For given $(\rho_c,\nu_c)$, one can integrate Eqs.~\eqref{eq:dX}--\eqref{eq:hydro-r} from the center to
the surface of the star, $r=R$, defined by $p(R)=0$.
At $r=R$, the internal solution is matched smoothly to the external solution described
by Eqs.~\eqref{eq:ext-soln-met} and~\eqref{eq:ext-soln-X}.
We require that
$\rho(R)=\rho'(R)=0$ and exclude the case where $\rho'$ is discontinuous across the surface.
For $X$ to satisfy $X(R)=1/2$,
$\nu_c$ must be adjusted to a suitable value.
The value of the integration constant $\mu$ is then determined from $e^{\nu(R)}=1-2\mu/R$.
One can thus obtain a family of unperturbed configurations parametrized by the single parameter $\rho_c$.
Note that suitable $\nu_c$ does not necessarily exit, depending on $\rho_c$.

\subsection{Weak-field limit}\label{subsec:weak-field}

In this paper, we will solve the field equations for the unperturbed background fully numerically
to obtain results valid even in the strong-field regime.
However, it is instructive to see analytically
the case of weak gravitational fields sourced by a nonrelativistic fluid body,
as our TOV system is greatly simplified.
To do so, let us assume that 
\begin{align}
        \nu=\delta\nu\ll 1,\quad X-\frac{1}{2}=\delta X\ll 1.
\end{align}
We also ignore the pressure $p$.
To first order in $\delta\nu$, $\delta X$, and $\rho$,
Eqs.~\eqref{eq:dX} and~\eqref{eq:dnu} reduce to 
\begin{align}
        \delta X'&=\frac{\delta\nu'+(\delta\nu+2\delta X)/r}{2(\alpha_H+\beta_1)}
        +\frac{r\rho}{4M^2(\alpha_H+2\beta_1)},\label{eq:deltaX}
        \\ 
        \delta\nu'&=-\frac{\delta\nu+2\delta X}{r}
        \notag \\ &\quad 
        -\frac{r(\alpha_H+\beta_1)[(1+3\alpha_H+3\beta_1)\rho+(\alpha_H+\beta_1)r\rho']}{2M^2(\alpha_H+2\beta_1)(1-\alpha_H-3\beta_1)}.
        \label{eq:deltanu}
\end{align}
Here we introduced the convenient parametrization~\cite{Langlois:2017mxy},
\begin{align}
        &M^2:=2f(1/2),\quad \alpha_H:=\left.-\frac{2Xf_X}{f}\right|_{X=1/2},
        \notag \\ & 
        \beta_1:=\left.\frac{X}{f}\left(f_X+XA_3\right)\right|_{X=1/2},
        \label{eq:def-eft}
\end{align}
which is conventionally used in the context of the effective field theory of dark energy and modified gravity.
Combining Eqs.~\eqref{eq:deltaX} and~\eqref{eq:deltanu},
we obtain
\begin{align}
        \left(r^2\delta\nu'\right)'=2G_N
        \left[
                \mathcal{M}+\gamma_1r^2\mathcal{M}''
        \right]',
\end{align}
where we defined
\begin{align}
        8\pi G_N&:=\frac{1}{M^2(1-\alpha_H-3\beta_1)},\label{def:GNewton}
        \\ 
        \gamma_1&:=-\frac{(\alpha_H+\beta_1)^2}{2(\alpha_H+2\beta_1)},
\end{align}
and 
\begin{align}
        \mathcal{M}(r):=4\pi\int^r_0 \rho(\tilde r) \tilde r^2\D\tilde r.
\end{align}
This immediately leads to 
\begin{align}
        \delta\nu'=2G_N\left(\frac{\mathcal{M}}{r^2}+\gamma_1\mathcal{M}''\right),
        \label{eq:v-b-newton}
\end{align}
which reproduces the Newtonian limit of
DHOST theories~\cite{Kobayashi:2014ida,Crisostomi:2017lbg,Langlois:2017dyl,Dima:2017pwp}.
The second term in the right-hand side of Eq.~\eqref{eq:v-b-newton} is nonvanishing only inside a fluid body,
and hence it tells us
how the breaking of Vainshtein screening occurs in the Newtonian regime.
The deviation from the standard Newtonian result is parametrized by $\alpha_H$ and $\beta_1$.

Using Eqs.~\eqref{eq:deltanu} and~\eqref{eq:v-b-newton}, we obtain 
\begin{align}
        \delta X&=-\frac{\delta\nu}{2}-\frac{G_N\mathcal{M}}{r}
        \notag \\ & \quad 
        -\frac{(\alpha_H+\beta_1)(1+\alpha_H+\beta_1)}{2(\alpha_H+2\beta_1)}G_N\mathcal{M}'.
        \label{eq:sonldx}
\end{align}
To first order in the small quantities, Eq.~\eqref{eq:elambda} yields
\begin{align}
        e^\lambda=1+r\left(\delta\nu'-2\alpha_H\delta X'\right).
\end{align}
Substituting Eqs.~\eqref{eq:v-b-newton} and~\eqref{eq:sonldx} to this,
we obtain
\begin{align}
        e^\lambda=1+2G_N\left(\frac{\mathcal{M}}{r}
        +\alpha_H\mathcal{M}'+\gamma_3 r\mathcal{M}''
        \right),
\end{align}
where 
\begin{align}
        \gamma_3:=-\frac{\beta_1(\alpha_H+\beta_1)}{2(\alpha_H+2\beta_1)}.
\end{align}
This also reproduces the previous result for the Vainshtein regime in the weak-field approximation~\cite{Kobayashi:2014ida,Crisostomi:2017lbg,Langlois:2017dyl,Dima:2017pwp}.
The deviation from standard gravity is parametrized again by $\alpha_H$ and $\beta_1$.

The observational constraints on $\alpha_H$ and $\beta_1$ are found to be~\cite{Dima:2017pwp}
\begin{align}
        -0.05<\alpha_H<0.26,\quad -0.08<\beta_1<0.02,
\end{align}
which come from the combination of the Hulse-Taylor pulsar~\cite{BeltranJimenez:2015sgd} and stellar structure physics~\cite{Saltas:2018mxc}.
(See Ref.~\cite{Saltas:2022ybg} for updated constraints from stellar physics,
though they are not given in terms of $\alpha_H$ and $\beta_1$ explicitly.)

\section{Odd-parity perturbations}\label{sec:Odd-pert}

For the calculation of the tidal Love numbers,
we consider the linear metric perturbations:
\begin{align}
        g_{\mu\nu}=\bar g_{\mu\nu}+h_{\mu\nu}.
\end{align}
In this paper, we only study the gravitomagnetic Love numbers,
which are obtained from the odd-parity sector of perturbations~\cite{Damour:2009vw,Binnington:2009bb,Landry:2015cva}.
Note that there is no odd-parity perturbation in the scalar field.
The nonvanishing components of the odd-parity metric perturbations in the Regge-Wheeler gauge are~\cite{Regge:1957td} 
\begin{align}
        h_{t\theta}&=
        -\frac{1}{\sin\theta}\sum_{\ell,m}h_0^{(\ell m)}(t,r)\partial_\varphi Y_{\ell m},
        \\ 
        h_{t\varphi}&=
        \sin\theta\sum_{\ell,m}h_0^{(\ell m)}(t,r)\partial_\theta Y_{\ell m},
        \\ 
        h_{r\theta}&=
        -\frac{1}{\sin\theta}\sum_{\ell,m}h_1^{(\ell m)}(t,r)\partial_\varphi Y_{\ell m},
        \\ 
        h_{r\varphi}&=
        \sin\theta\sum_{\ell,m}h_1^{(\ell m)}(t,r)\partial_\theta Y_{\ell m},
\end{align}
with $Y_{\ell m}(\theta,\varphi)$ being the spherical harmonics.
It is straightforward to compute the nonvanishing components of the perturbed $\mathcal{E}_\mu^\nu$:
\begin{widetext}
\begin{align}
        \delta\mathcal{E}_\varphi^t&=\sum_{\ell,m}
        \left\{\frac{e^{-\nu/2-\lambda/2}}{r^2}
        \left[
                r^2e^{-\nu/2-\lambda/2}f\left(h_0'-\frac{2}{r}h_0-\dot h_1\right)
        \right]'
        -\frac{(\ell-1)(\ell+2)}{r^2}e^{-\nu}fh_0\right\}
        \sin\theta\partial_\theta Y_{\ell m},
        \\ 
        \delta\mathcal{E}_\varphi^r&=\sum_{\ell,m}
        e^{-\nu-\lambda}f\left[
                \ddot h_1-\dot h_0'+\frac{2}{r}\dot h_0
                +\frac{(\ell-1)(\ell+2)}{r^2}e^{\nu}h_1
        \right]\sin\theta\partial_\theta Y_{\ell m},
        \\ 
        \delta \mathcal{E}_\varphi^\theta&=\sum_{\ell,m}
        \left\{
        -\frac{e^{-\nu/2-\lambda/2}}{r^2}\left[
                e^{\nu/2-\lambda/2}fh_1
        \right]'+\frac{e^{-\nu}}{r^2}f\dot h_0\right\}
        \left(
                \cos\theta\partial_\theta Y_{\ell m}-\sin\theta 
                \partial_\theta \partial_\theta Y_{\ell m}
        \right),
\end{align}
\end{widetext}
where a dot stands for differentiation with respect to $t$.
Here and hereafter, we omit the labels $(\ell m)$ for notational simplicity.

The odd-parity sector of the fluid perturbations arises from the perturbed four-velocity,
\begin{align}
        \delta u^\mu=\sum_{\ell,m}
        \frac{e^{-\nu/2}U(t,r)}{r^2(\rho+p)}\left(
                0,0,\frac{\partial_\varphi Y_{\ell m}}{\sin\theta}
                ,\frac{\partial_\theta Y_{\ell m}}{\sin\theta}
        \right).
\end{align}
As clarified in Ref.~\cite{Pani:2018inf}, there are two approaches to the calculation of the gravitomagnetic Love numbers depending on the assumptions made on the fluid perturbations.
In this paper, we follow the irrotational fluid approach of~\cite{Damour:2009vw,Landry:2015cva},
in which the zero-frequency limit of the Regge-Wheeler equation is taken at the end
instead of setting $U=0$ from the beginning.
The perturbed energy-momentum tensor is then given by
\begin{align}
        \delta T_\varphi^t&=\sum_{\ell,m}e^{-\nu}\left[
                (\rho+p)h_0+U
        \right]\sin\theta\partial_\theta Y_{\ell m},
        \label{eq:comp-dT}
        \\  
        \delta T_\varphi^r&=\delta T_\varphi^\theta=0.
\end{align}
The $\varphi$-component of the hydrodynamical equations gives
\begin{align}
        \delta\left(\nabla_\mu T_\varphi^\mu \right)
        = \partial_t\delta T_\varphi^t=0.
        \label{eq:hydro-pert-varphi}
\end{align}

We assume the time dependence of the perturbations as
$h_0(t,r)=h_0(r)e^{-i\omega t}$,
$h_1(t,r)=h_1(r)e^{-i\omega t}$, and
$U(t,r)=U(r)e^{-i\omega t}$.
It follows from Eqs.~\eqref{eq:comp-dT} and~\eqref{eq:hydro-pert-varphi} that 
\begin{align}
        -i\omega\left[(\rho+p)h_0+U\right]=0.\label{eq:hydro-Fou}
\end{align}

To derive a single master equation for the odd-parity perturbations,
it is convenient to introduce~\cite{Takahashi:2019oxz,Tomikawa:2021pca}
\begin{align}
        \chi(r):=f\left(h_0'-\frac{2}{r}h_0+i\omega h_1\right).
        \label{eq:def-chi}
\end{align}
Using this variable, one can write the perturbed gravitational-field equations
$\delta \mathcal{E}_\mu^\nu=\delta T_\mu^\nu$ as 
\begin{align}
        &e^{\nu/2-\lambda/2}\left(
                r^2e^{-\nu/2-\lambda/2}\chi
        \right)'-(\ell-1)(\ell+2)fh_0
        \notag \\ 
        &=r^2\left[(\rho+p)h_0+U\right]=0,
        \label{eq:pert1}
        \\ 
        &i\omega\chi+\frac{(\ell-1)(\ell+2)}{r^2}e^\nu f h_1=0,
        \label{eq:pert2}
        \\ 
        &e^{\nu/2-\lambda/2}\left[
                e^{\nu/2-\lambda/2}fh_1
        \right]'+i\omega f h_0=0.
        \label{eq:pert3}
\end{align}
Using Eqs.~\eqref{eq:pert1} and~\eqref{eq:pert2},
we can remove $h_0$ and $h_1$ from Eq.~\eqref{eq:def-chi} to get 
\begin{align}
        &
        e^{-\nu}\omega^2\chi+e^{-\lambda}\chi''
        -e^{-\lambda}\frac{F'}{F}\chi'
        \notag \\ &
        -
        \left\{
                \frac{\ell(\ell+1)-2}{r^2}+f\left[
                        \frac{1}{F}\left(
                                \frac{F}{f}
                        \right)'
                \right]'
        \right\}\chi=0,\label{eq:master-eq}
\end{align}
with $F(r):=e^{\nu/2+3\lambda/2}f/r^2$.
This is the master equation for the odd-parity perturbations.
One can reconstruct $h_0$, $h_1$, and $U$ from $\chi$
with the help of Eqs.~\eqref{eq:hydro-Fou},~\eqref{eq:pert1}, and~\eqref{eq:pert2}.

We are interested in the solutions in the zero-frequency limit, $\omega\to 0$.
Rather than solving the master equation~\eqref{eq:master-eq}, we derive
a differential equation for $h_0$, which is more directly related to the Love number calculation.
Using Eqs.~\eqref{eq:pert1} and~\eqref{eq:master-eq},
it can be shown that $h_0$ in the zero-frequency limit obeys the equation 
\begin{align}
        h_0''-\frac{\mathcal{P}(r)}{r}h_0'-\frac{\mathcal{Q}(r)}{r^2}h_0=0,
        \label{eq:master-h0}
\end{align}
where 
\begin{align}
        \mathcal{P}(r)&:=r\left[\frac{\nu'}{2}+\frac{\lambda'}{2}-\frac{f_XX'}{f}\right],
        \label{eq:def-calP}
        \\ 
        \mathcal{Q}(r)&:=e^\lambda\left[\ell(\ell+1)-2\left(1-e^{-\lambda}\right)
        \right]
        -2\mathcal{P}(r).
        \label{eq:def-calQ}
\end{align}
We can see that the effects of modified gravity appear in two ways.
First, the modification of the unperturbed configuration appears through $\nu$ and $\lambda$.
Second, the new term $f_XX'/f$ modifies directly the form of the equation for $h_0$ as compared to the corresponding equation in GR.

In the external region where the unperturbed solution is given by Eqs.~\eqref{eq:ext-soln-met} and~\eqref{eq:ext-soln-X},
Eq.~\eqref{eq:master-h0} reduces to the same equation as in GR,
leading to the same analytic solution~\cite{Landry:2015cva}
\begin{align}
        h_0=\frac{2}{3(\ell-1)}r^{\ell+1}\left[
                A_\ell(r)-4\frac{\ell+1}{\ell}\tilde k_\ell R^{2\ell} \mu\frac{B_\ell(r)}{r^{2\ell+1}}
        \right],
\end{align}
where 
\begin{align}
        A_\ell(r)&:={}_2F_1(-\ell+1,-\ell-2,-2\ell, 2\mu/r),
        \\ 
        B_\ell(r)&:={}_2F_1(\ell-1,\ell+2,2\ell+2,2\mu/r),
\end{align}
with ${}_2F_1$ being the hypergeometric function. 
Note that $A_\ell(r),B_\ell(r)=1+\mathcal{O}(\mu/r)$ for $r\gg \mu$.
The dimensionless coefficient $\tilde k_\ell$ is the (rescaled) gravitomagnetic Love number~\cite{Landry:2015cva}.
By requiring that $h_0$ and $h_0'$ are continuous across the stellar surface $r=R$,
we obtain
\begin{align}
        \tilde k_\ell=\left.
        \frac{\ell}{4(\ell+1)}\frac{R}{\mu}
        \frac{RA'_\ell-(\kappa-\ell-1)A_\ell}{RB'_\ell-(\kappa+\ell)B_\ell}
        \right|_{r=R},
\end{align}
where $\kappa(r):=rh_0'/h_0$.
For the calculation of $\tilde k_\ell$, it is therefore convenient to rewrite Eq.~\eqref{eq:master-h0} in terms of $\kappa$ as
\begin{align}
        r\kappa'+\kappa^2-(1+\mathcal{P})\kappa-\mathcal{Q}=0.
        \label{eq:master-eta}
\end{align}
In the vicinity of the center, $\nu$, $\lambda$, and $X$ are expanded as 
Eqs.~\eqref{bc:1},~\eqref{bc:2}, and~\eqref{bc:4},
and hence $\mathcal{P}=\mathcal{O}(r^2)$ and $\mathcal{Q}=\ell(\ell+1)+\mathcal{O}(r^2)$.
This shows that the boundary condition is given by $h_0\propto r^{\ell +1}$ at $r=0$,
which is translated to
\begin{align}
        \kappa(0)=\ell+1.
        \label{eq:bc-eta}
\end{align}
Equation~\eqref{eq:master-eta} can be integrated numerically from $r=0$ to $r=R$
with the boundary condition~\eqref{eq:bc-eta}.
Since the internal solution for $\kappa(r)$ differs from that in GR due to the modified interior structure and nonvanishing $X'$,
the Love numbers $\tilde k_\ell$ will also be different from the values in GR,
signaling the effects of the partial breaking of Vainshtein screening.

\section{Model description}\label{sec:model}

\begin{figure*}[t]
        \centering
        \begin{minipage}[b]{1.02\columnwidth}
            \centering
            \includegraphics[width=1.0\columnwidth]{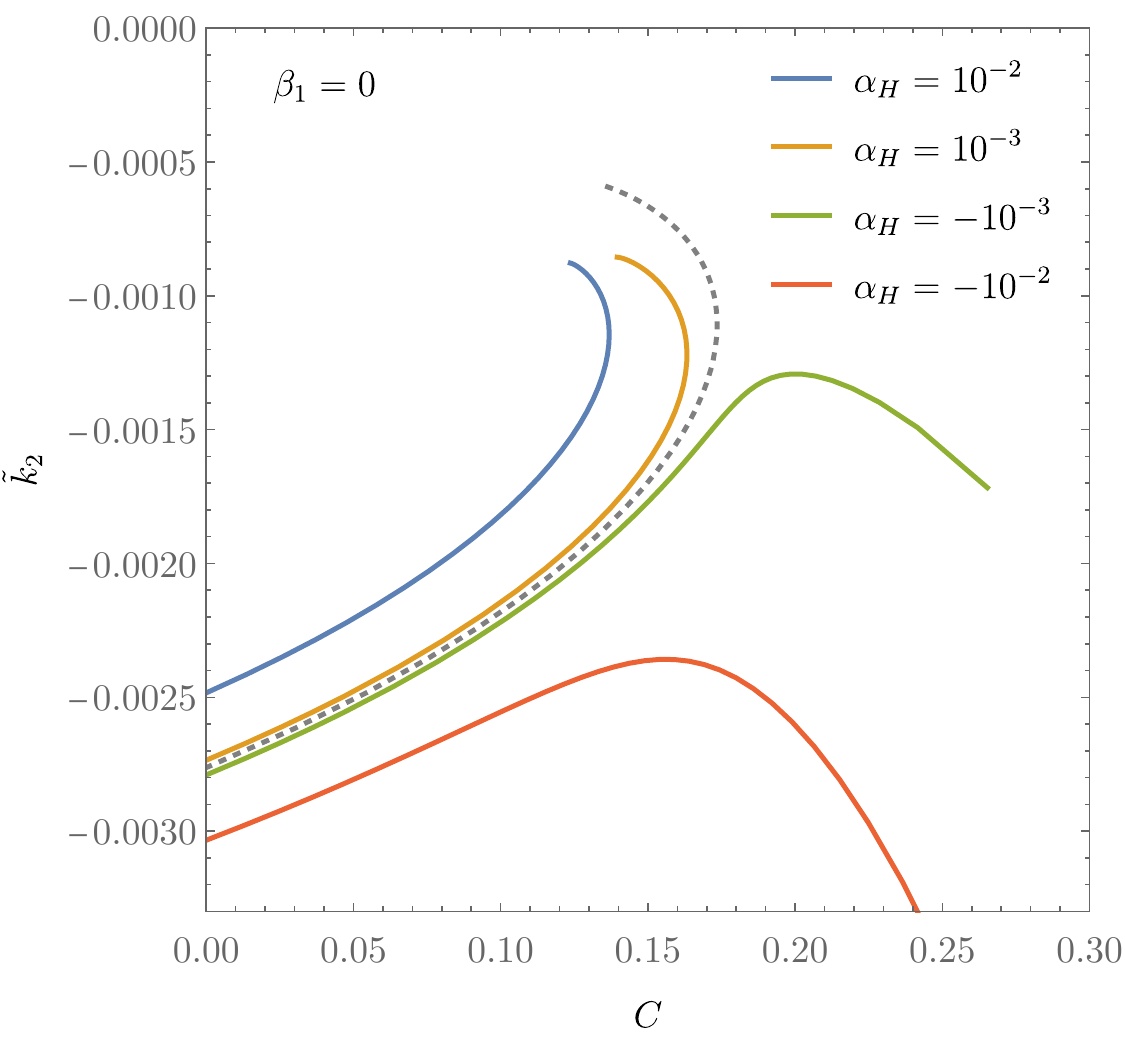}
        \end{minipage}
        \begin{minipage}[b]{0.9\columnwidth}
            \centering
            \includegraphics[width=0.86\columnwidth]{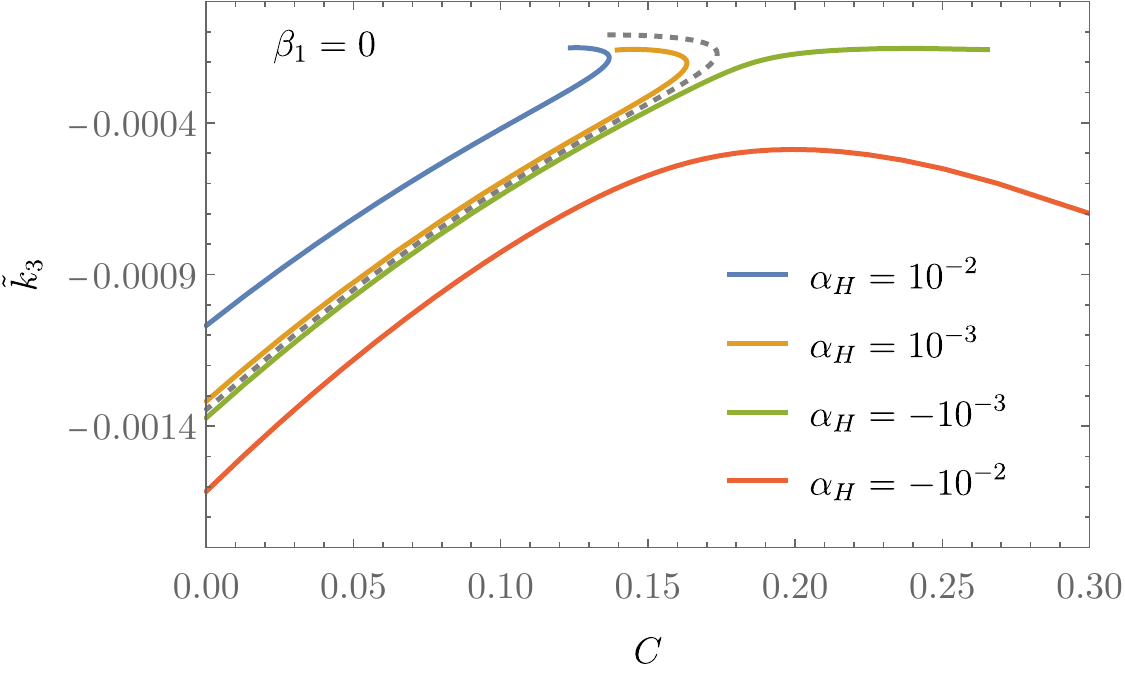}
            \includegraphics[width=0.88\columnwidth]{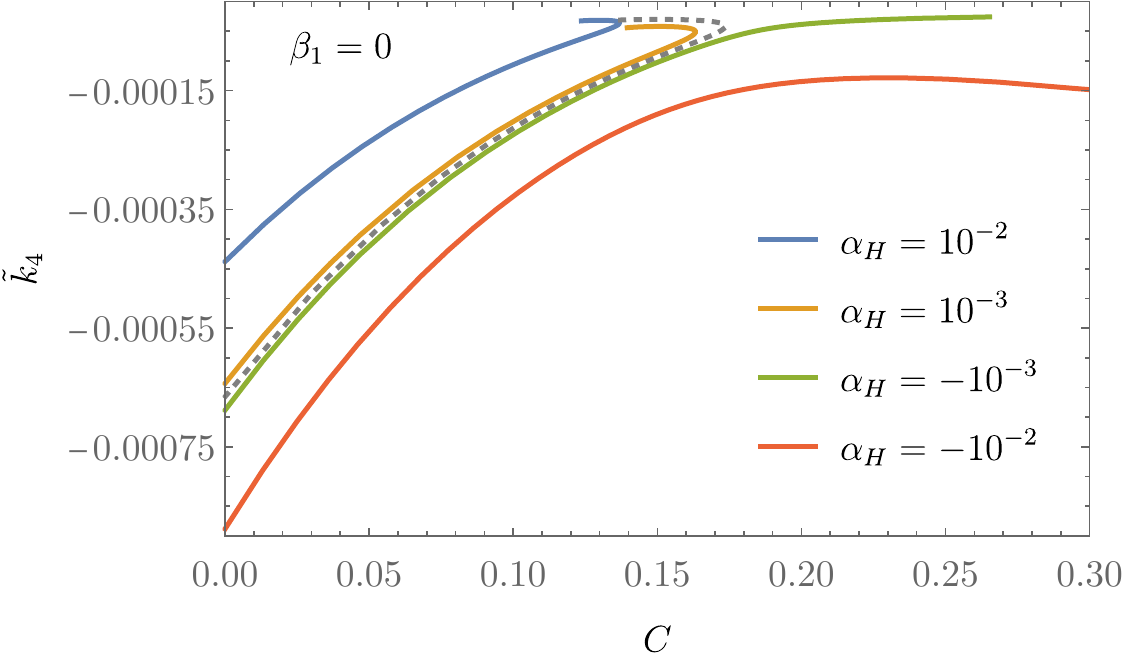}
        \end{minipage}
        \caption{Gravitomagnetic Love numbers $\tilde k_2$ (left), $\tilde k_3$ (upper right), and $\tilde k_4$ (lower right)
        for different values of $\alpha_H$ versus compactness $C=\mu/R$.
        The other model parameters are given by $\beta_1=0$ and $q=1$. We use the polytropic index $n=2$.
        The dashed lines represent the results for GR.}
        \label{fig:n2-k234}
\end{figure*}

\begin{figure*}[t]
        \centering
        \begin{minipage}[b]{0.98\columnwidth}
            \centering
            \includegraphics[width=0.9\columnwidth]{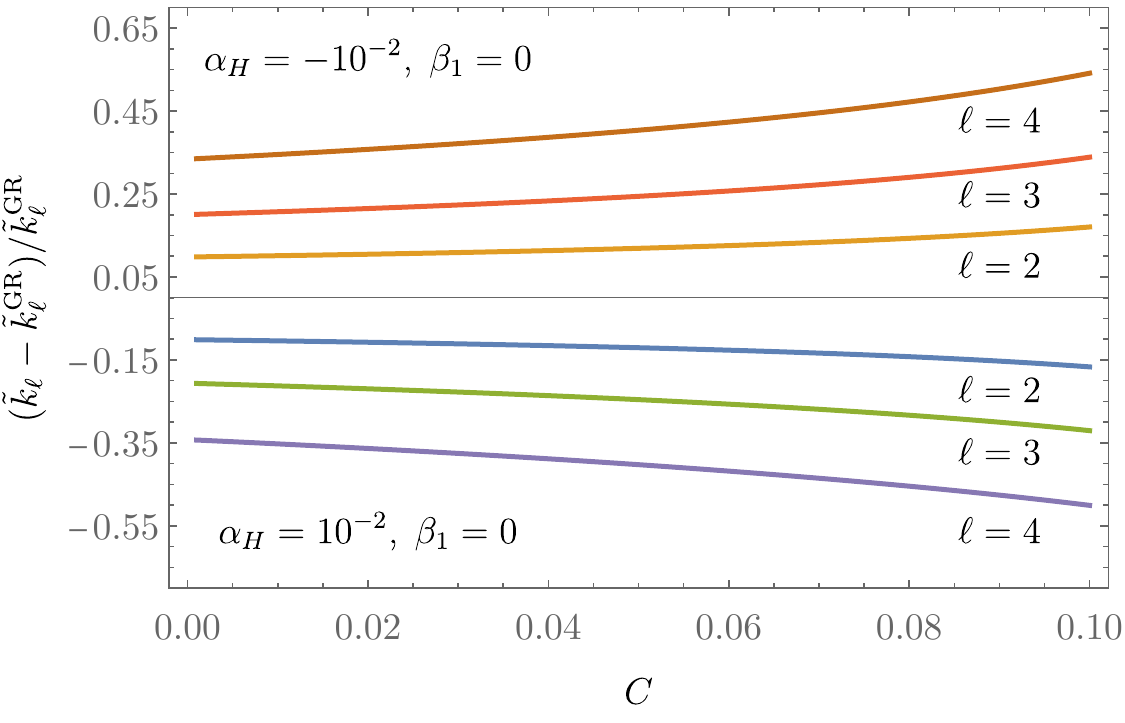}
        \end{minipage}
        \begin{minipage}[b]{0.98\columnwidth}
            \centering
            \includegraphics[width=0.9\columnwidth]{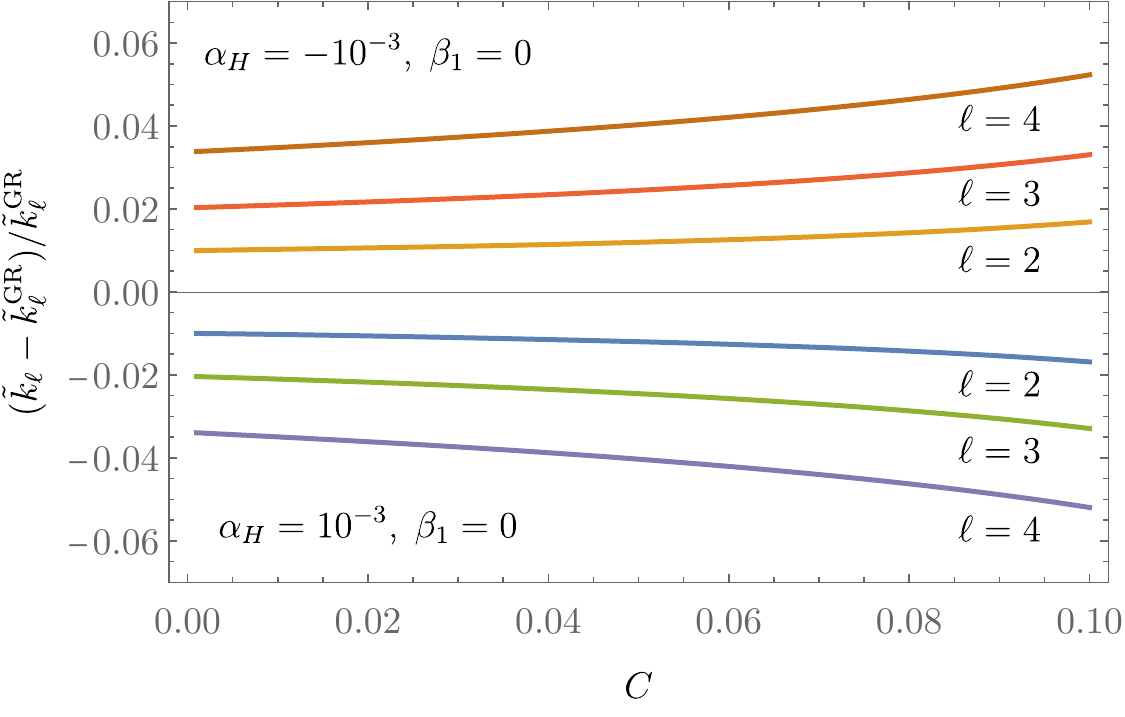}
        \end{minipage}
        \caption{Relative differences between the $\ell=2,3,4$
        gravitomagnetic Love numbers in DHOST theories and GR versus compactness.
        The left panel shows the results for $\alpha_H=\pm 10^{-2}$, while the right panel is
        for $\alpha_H=\pm 10^{-3}$.
        The other model parameters are given by $\beta_1=0$ and $q=1$. We use the polytropic index $n=2$.}
        \label{fig:ratio-2-3}
\end{figure*}

\begin{figure*}[t]
        \centering
        \begin{minipage}[b]{1.02\columnwidth}
            \centering
            \includegraphics[width=1.0\columnwidth]{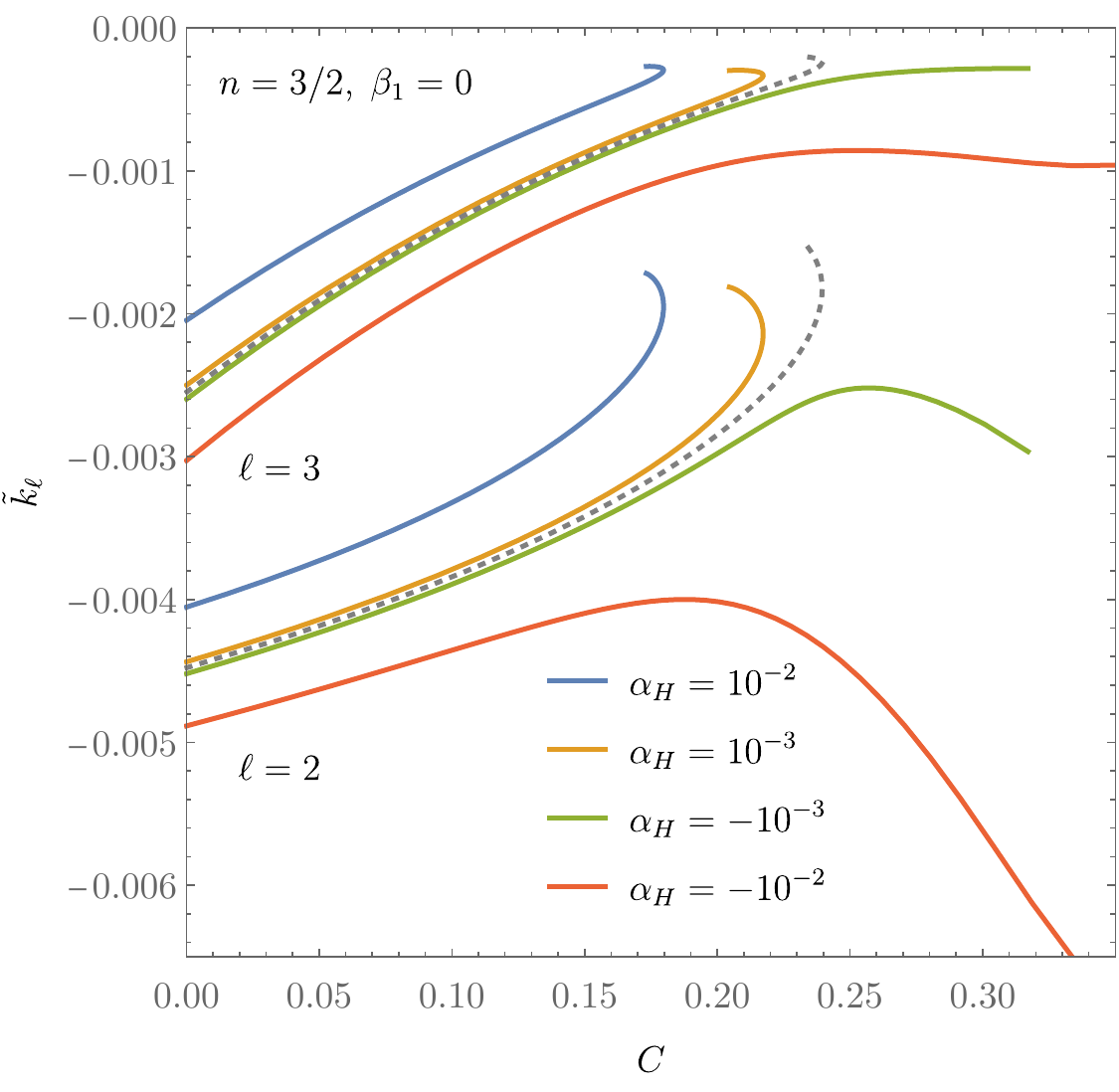}
        \end{minipage}
        \begin{minipage}[b]{0.9\columnwidth}
            \centering
            \includegraphics[width=0.86\columnwidth]{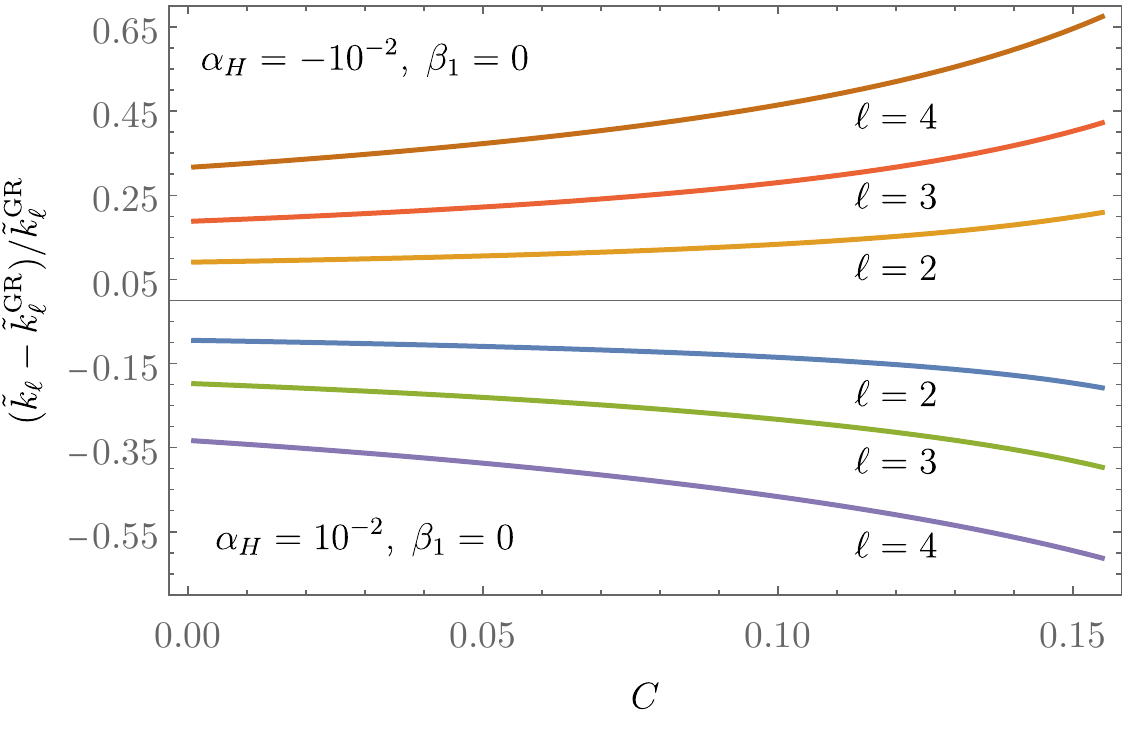}
            \includegraphics[width=0.88\columnwidth]{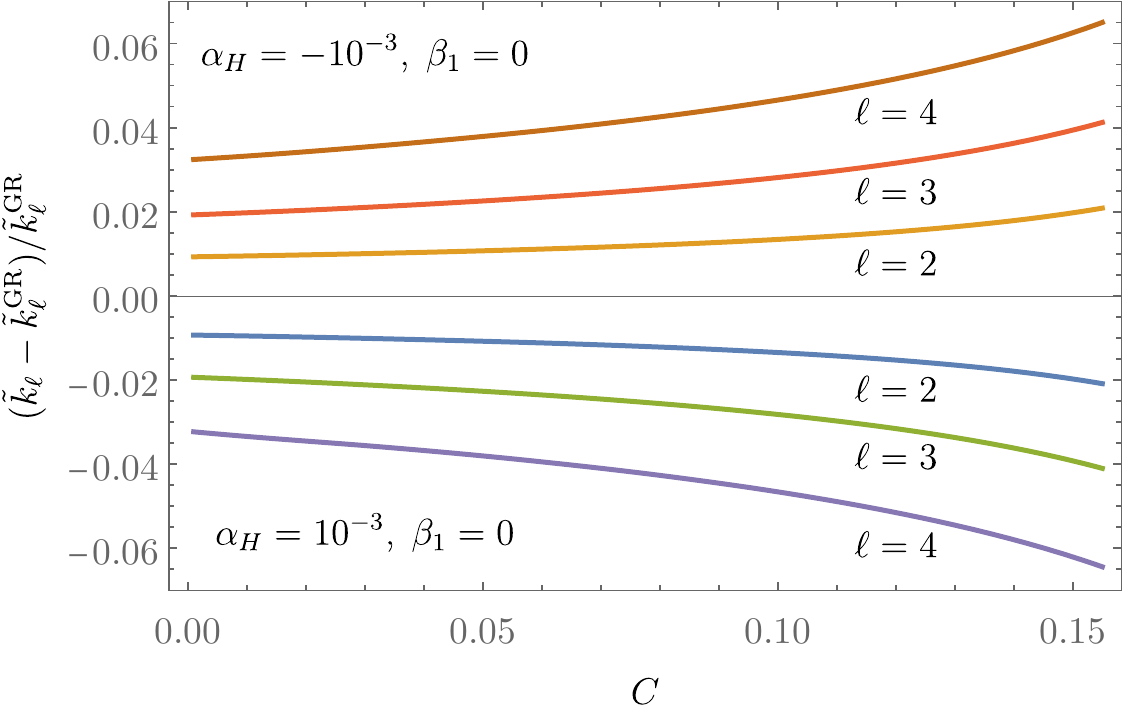}
        \end{minipage}
        \caption{Gravitomagnetic Love numbers in DHOST theories versus compactness (left)
        and their relative differences compared to GR (upper right and lower right).
        We use the polytropic index $n=3/2$.
        The dashed lines in the left panel represent the results for GR.}
        \label{fig:n32k23.pdf}
\end{figure*}

\begin{figure}[tb]
        \begin{center}
        \includegraphics[width=0.9\columnwidth]{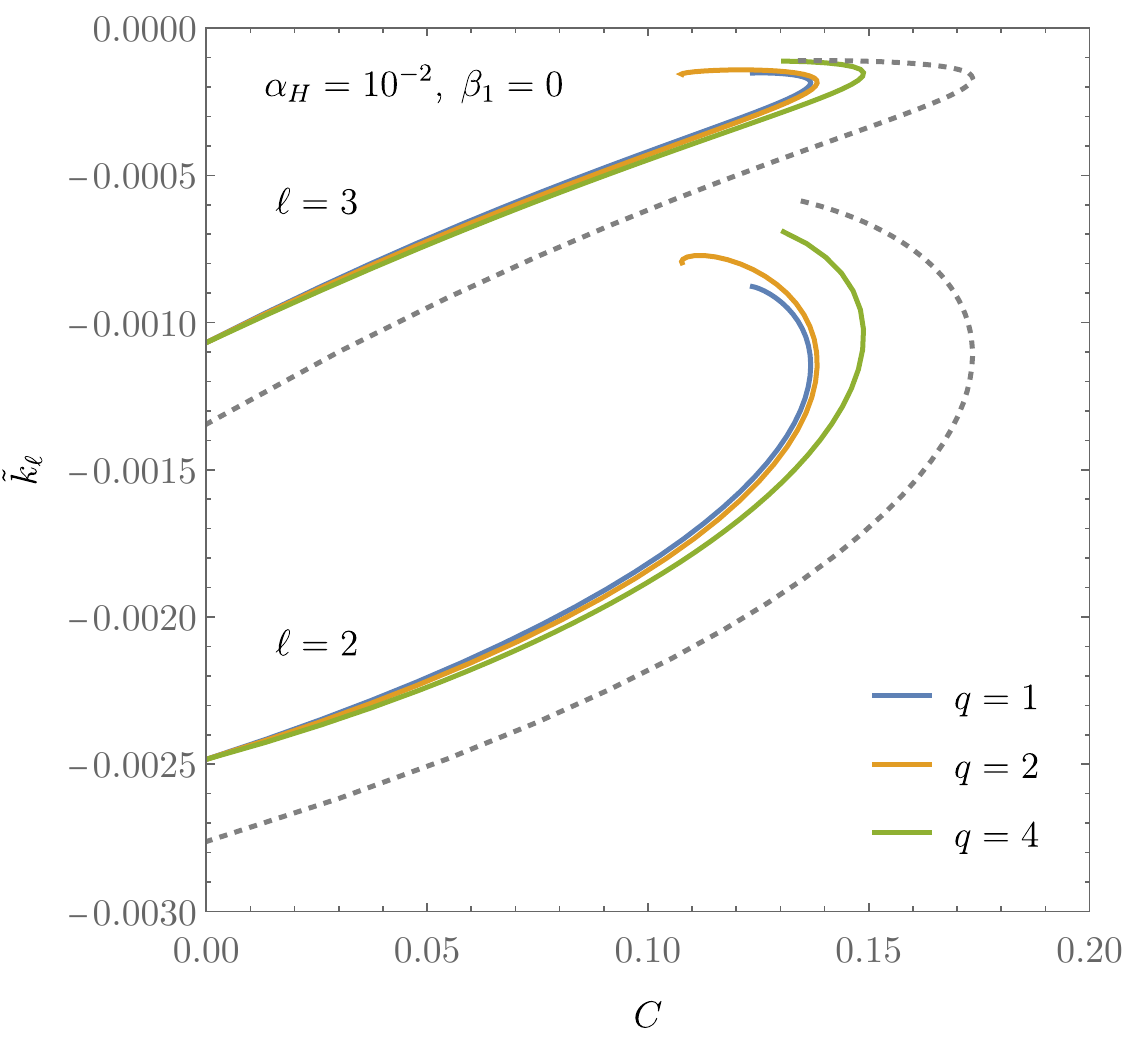}
        \end{center}
        \caption{Gravitomagnetic Love numbers in DHOST theories with $\alpha_H=10^{-2}$ and $\beta_1=0$
        versus compactness. We compare the results for different values of $q$. We use the polytropic index $n=2$.
        The dashed lines represent the results for GR.}
        \label{fig:q124.pdf}
\end{figure}

\begin{figure*}[t]
        \centering
        \begin{minipage}[b]{0.98\columnwidth}
            \centering
            \includegraphics[width=0.93\columnwidth]{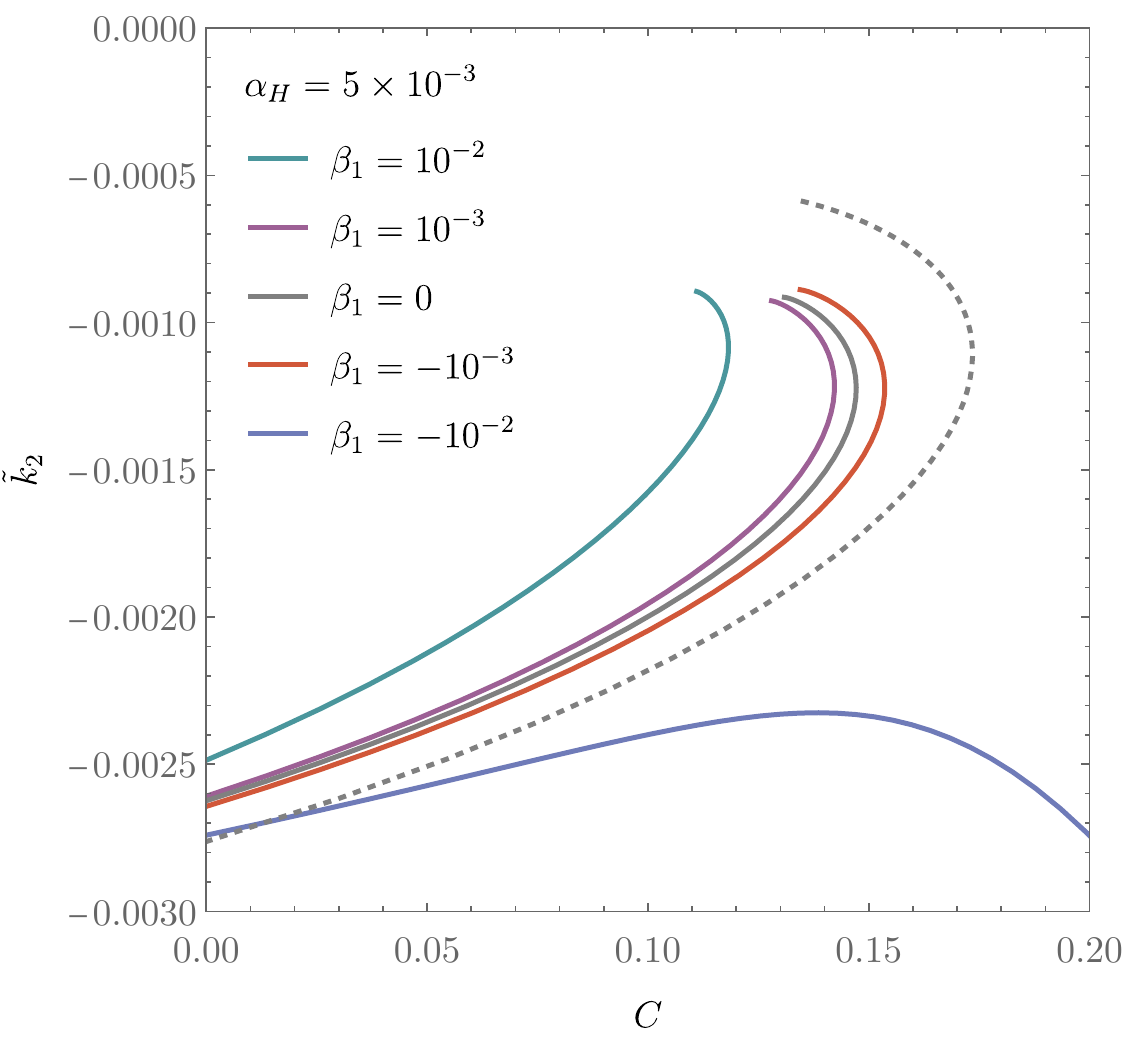}
        \end{minipage}
        \begin{minipage}[b]{0.98\columnwidth}
            \centering
            \includegraphics[width=0.9\columnwidth]{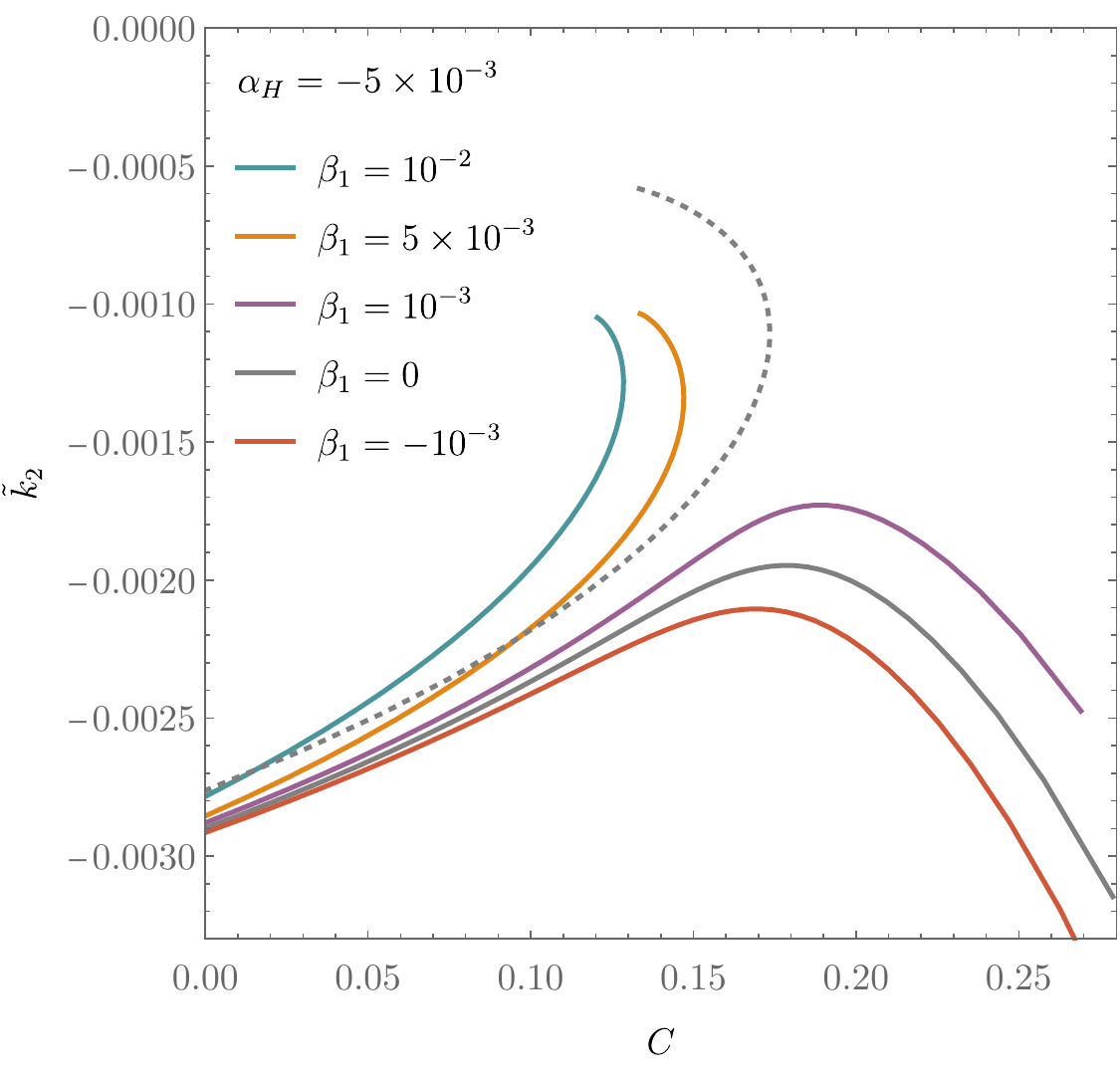}
        \end{minipage}
        \caption{$\ell=2$ gravitomagnetic Love number $\tilde k_2$ for different values of $\beta_1$
        versus compactness.
        The left panel shows the results for $\alpha_H=5\times 10^{-3}$, while
        the right panel is for $\alpha_H=-5\times 10^{-3}$.
        We use $q=1$ and the polytropic index $n=2$.
        The dashed lines represent the results for GR.}
        \label{fig:beta-p-m}
\end{figure*}

\subsection{Modified gravity model}

So far we have derived the general equations governing the unperturbed TOV system
and the odd-parity perturbations in DHOST theories.
To proceed,
we consider a family of concrete DHOST models with the functions of the form
\begin{align}
        f=f_0+f_1X^q,\quad A_3=a_3,
\end{align}
where $f_0$, $f_1$, $a_3$, and $q$ are constant parameters.
Using Eq.~\eqref{eq:def-eft},
one can express $f_0$, $f_1$, and $a_3$ in terms of $M^2$, $\alpha_H$, and $\beta_1$ to obtain
the following useful expression:
\begin{align}
        f&=\frac{M^2}{2}\left[
                1+\frac{\alpha_H}{2q}-\frac{\alpha_H}{2q}(2X)^{q}
        \right],
        \label{eq:concrete-f}
        \\
        A_3&=\frac{M^2}{4}(\alpha_H+2\beta_1).
        \label{eq:concrete-A3}
\end{align}
The deviations from the standard behavior of gravity
in the weak-field regime are parametrized completely by $\alpha_H$ and $\beta_1$,
as we have seen in Sec.~\ref{subsec:weak-field}.
However, they alone will be insufficient for the characterization of the strong-field regime
and the effects of $q$ are expected to emerge when gravity gets stronger.
The model with $q=2$ was studied in Refs.~\cite{Crisostomi:2017pjs,Kobayashi:2018xvr,Hiramatsu:2020fcd,Hiramatsu:2022fgn}.
By comparing the results with the same $(\alpha_H,\beta_1)$ but with different $q$,
one would be able to disclose the effects that are not captured by
the conventionally used parameters $(\alpha_H,\beta_1)$.

We use the DHOST theory with the functions~\eqref{eq:concrete-f} and~\eqref{eq:concrete-A3}
to compute the tidal Love numbers.

\subsection{Relativistic star model}

We are not attempting to determine precisely the tidal Love numbers in realistic situations.
We therefore compute the Love numbers for energy polytropes, which have the simple equation of state
\begin{align}
        p=K\rho^{1+1/n}.
\end{align}
The energy polytropes with $n\le 1$ result in $\rho'(R)\neq 0$,
which hinders a smooth matching of the internal and external metrics at the surface of the star
in DHOST theories.
Though the form of the equation of state is different,
discontinuities essentially caused by this are found in the previous study~\cite{Kobayashi:2018xvr}.
To avoid such discontinuities, we restrict our investigation to the polytropic indices
$n>1$ (more specifically, $n=2$ and $n=3/2$).

It is convenient to introduce the parameter
\begin{align}
        b:=K\rho_c^{1/n},
\end{align}
which is a dimensionless measure of the central density.
For a given set of the model parameters $(\alpha_H,\beta_1,q)$ and the polytropic index $n$, 
Eqs.~\eqref{eq:dX}--\eqref{eq:hydro-r} can be integrated to
yield a sequence of unperturbed background solutions parametrized by $b$.
In the actual numerical computations, we use the dimensionless energy density
$\Theta:=(\rho/\rho_c)^{1/n}$ and the dimensionless radial coordinate
$\tilde r:=(r/M)\sqrt{\rho_c/b}$, which allows us to rewrite the field equations in a dimensionless form. The resultant field equations are solved using NDSolve of \textit{Mathematica}.

We find that there is a maximum value $b=b_{\textrm{max}}$ above which
one cannot find $\nu_c$ that admits an appropriate matching of the internal and external solutions.
The same absence of a solution above a certain central density in DHOST theories
was reported in Ref.~\cite{Kobayashi:2018xvr} using a different equation of state.
A similar result was also found in Ref.~\cite{Cisterna:2015yla}
in a subset of the Horndeski theory with the coupling between the Einstein tensor and
the first derivative of the scalar field.
We also find the following qualitative results for the unperturbed configurations.
For fixed $q$ and $\beta_1$, positive (negative) $\alpha_H$ makes the maximum compactness
smaller (larger).
Similarly, for fixed $q$ and $\alpha_H$, positive (negative) $\beta_1$ makes the maximum compactness
smaller (larger).
Here, the compactness is defined as
\begin{align}
        C:=\frac{\mu}{R}.
\end{align}

\section{Love numbers}\label{sec:Love}

The gravitomagnetic Love numbers $\tilde k_\ell$ are computed numerically for different model parameters
as functions of the compactness $C$, using NDSolve of \textit{Mathematica} to solve Eq.~\eqref{eq:master-eta}.
Our results are displayed in Figs.~\ref{fig:n2-k234}--\ref{fig:beta-p-m}.

Figure~\ref{fig:n2-k234} shows the Love numbers $\tilde k_2$, $\tilde k_3$, and $\tilde k_4$ for
different values of $\alpha_H$.
The other model parameters are fixed as $\beta_1=0$ and $q=1$, and
the polytropic index is given by $n=2$.
We present in Fig.~\ref{fig:ratio-2-3} the relative difference,
$(\tilde k_\ell-\tilde k_\ell^{\textrm{GR}})/\tilde k_\ell^{\textrm{GR}}$,
where $\tilde k_\ell^{\textrm{GR}}$ is the Love number in GR.
Typically, we have 
\begin{align}
        \frac{\tilde k_\ell-\tilde k_\ell^{\textrm{GR}}}{\tilde k_\ell^{\textrm{GR}}}
        \sim -\mathcal{O}(10)\times \alpha_H,
\end{align}
and it is larger by some factor for higher multipoles.
We also find that the difference increases with larger compactness.

For the different choice of the polytropic index, $n=3/2$,
the Love numbers and their relative differences compared to GR are plotted in Fig.~\ref{fig:n32k23.pdf}. 
We find qualitatively the same results as those for $n=2$.

To see the effects that are not encapsulated solely in $\alpha_H$ and $\beta_1$,
we present in Fig.~\ref{fig:q124.pdf}
the Love numbers for different $q$, with $\alpha_H$ and $\beta_1$ being fixed as
$\alpha_H=10^{-2}$ and $\beta_1=0$.
For small $C$, the Love numbers depend almost only on $\alpha_H$ and $\beta_1$.
However, the $q$ dependence emerges as $C$ gets larger and one goes away from the weak-field regime.
This shows that $\alpha_H$ and $\beta_1$ alone are insufficient to capture the behavior of
gravity in the strong-field regime.

Figure~\ref{fig:beta-p-m} shows how the Love numbers depend on $\beta_1$ for selected values of $\alpha_H$.
We find that the two parameters $\alpha_H$ and $\beta_1$ have similar effects on
the tidal response:
positive (negative) $\alpha_H$ and $\beta_1$ contribute to
decreasing (increasing) the value of $|\tilde k_\ell|$.
The magnitude of the change in $\tilde k_\ell$ caused by $\beta_1$ is of the same
order as that by $\alpha_H$.

\section{Conclusions}\label{sec:conclusions}

We have studied the gravitomagnetic tidal deformability of relativistic stars in
scalar-tensor theories beyond Horndeski, i.e. degenerate higher-order scalar-tensor (DHOST) theories.
Though the Vainshtein mechanism operates completely outside,
screening is only partially effective in a region filled with matter in DHOST
theories~\cite{Kobayashi:2014ida,Crisostomi:2017lbg,Langlois:2017dyl,Dima:2017pwp}.
We have shown that this partial breaking of Vainshtein screening in astrophysical bodies
can be probed by their tidal response.

We have derived the odd-parity perturbation equations for the Love number calculation
in the zero-frequency limit, adopting the irrotational fluid approach of~\cite{Damour:2009vw,Landry:2015cva}.
The key equation is given by Eq.~\eqref{eq:master-h0},
in conjunction with the definitions~\eqref{eq:def-calP} and~\eqref{eq:def-calQ},
where Vainshtein-breaking effects arise from the nonvanishing gradient of
the scalar-field kinetic term as well as from the modified internal structure of the fluid body.
We numerically calculated the gravitomagnetic Love numbers, $\tilde k_\ell$,
for different parameters of the DHOST model defined
by the functions~\eqref{eq:concrete-f} and~\eqref{eq:concrete-A3}.
Two of the model parameters, $\alpha_H$ and $\beta_1$,
are conventionally used in the context of effective field theory of dark energy
and modified gravity
to characterize deviations from GR in the weak-field regime and linear cosmology.
We have found that the relative differences of the Love numbers compared to GR are
roughly $\sim \mathcal{O}(10)\times \alpha_H,\beta_1$.
The differences increase by some factor with higher multipoles and larger compactness.
Qualitatively, positive (negative) $\alpha_H$ and $\beta_1$ contribute to
decreasing (increasing) the value of $|\tilde k_\ell|$.
These two parameters are, however, insufficient to fully characterize
the deviations from GR away from the weak-field limit,
which we have demonstrated by changing the third parameter of the model, $q$.
Indeed, the effect of this parameter is more clearly seen for larger compactness.

In this paper, we have focused on the gravitomagnetic (odd-parity) Love numbers and postponed
the analysis of the gravitoelectric (even-parity) ones,
even though the latter have a larger contribution than the former to the phase
of the gravitational-wave signal from a binary inspiral.
The reason for this limitation is mostly technical.
The even-parity sector of the linear perturbations of a spherically symmetric solution
in DHOST theories is even more involved than that in the Horndeski theory.
The reduction of the number of derivatives in the degenerate system and the derivation of master variables have been successfully achieved
only for a specific background solution in vacuum DHOST theories~\cite{Takahashi:2021bml},
and at this moment it seems challenging to extend the procedure of~\cite{Takahashi:2021bml}
to the present setup for the Love number calculation.
We hope to come back to this issue in the future.

\acknowledgments
The work of TK was supported by
JSPS KAKENHI Grant No.~JP20K03936 and
MEXT-JSPS Grant-in-Aid for Transformative Research Areas (A) ``Extreme Universe'',
No.~JP21H05182 and No.~JP21H05189.

\appendix 

\section{Derivation of the equations in Sec.~\ref{subsec:TOV-system}}
\label{app:derivation-TOV}

In this appendix, we outline the derivation of the basic equation
governing the TOV system in DHOST theories,
following closely the previous work~\cite{Kobayashi:2018xvr}.

The first step is to notice that
$\mathcal{J}^r\propto \mathcal{E}_{tr}$,
and hence the gravitational field equation $\mathcal{E}_{tr}=0$ implies 
\begin{align}
        \mathcal{J}^r=0.
\end{align}
The scalar-field equation~\eqref{eom:scalar} is then automatically satisfied.

Using $X=[e^{-\nu}-e^{-\lambda}(\psi')^2]/2$ in place of $\psi'$,
we see, from explicit calculations, that the field equations are of the form 
\begin{align}
        \mathcal{E}_t^t&=b_1\nu''+b_2X''+b_3\lambda'+\tilde{\mathcal{E}}_t
        (\nu,\nu',\lambda,X,X')
        \notag \\ &=-\rho,\label{eom:tt}
        \\ 
        \mathcal{E}_r^r&=c_1\nu''+c_2X''+c_3\lambda'+\tilde{\mathcal{E}}_r
        (\nu,\nu',\lambda,X,X')
        \notag \\ &=p,\label{eom:rr}
        \\ 
        \psi'\mathcal{J}^r&=c_1\nu''+c_2X''+c_3\lambda'+\tilde{\mathcal{J}}
        (\nu,\nu',\lambda,X,X')
        \notag \\ &=0,\label{eom:J}
\end{align}
where $b_1$, $b_2$, $c_1$, $c_2$ are expressed in terms of $\nu$, $\lambda$, and $X$,
while $b_3$ and $c_3$ depend also on $\nu'$ and $X'$.
The point is that $\mathcal{E}_r^r$ and $\psi'\mathcal{J}^r$ share the same coefficients $c_1$, $c_2$, and $c_3$.
This fact allows us to solve the equation $\mathcal{E}_r^r-\psi'\mathcal{J}^r=p$ for $\lambda$ to obtain
\begin{align}
        e^\lambda=\mathcal{F}_\lambda(\nu,\nu',X,X',p).\label{eq-app:soln-lambda}
\end{align}

Equation~\eqref{eq-app:soln-lambda} can be used to remove $\lambda$ and $\lambda'$ from Eq.~\eqref{eom:J},
yielding 
\begin{align}
        \psi'\mathcal{J}^r=\eta_1\nu''+\eta_2X''+\tilde{\mathcal{E}}_1(\nu,\nu',X,X',\rho,p)=0,
        \label{eq-app:EE1}
\end{align}
where $\eta_{1,2}=\eta_{1,2}(\nu,\nu',X,X',p)$.
Here we used the hydrodynamical equation~\eqref{eq:hydro-r} to replace $p'$ with $\rho$, $p$, and $\nu'$.
Similar manipulations show that Eq.~\eqref{eom:tt} can also be rewritten in the form 
\begin{align}
        \eta_1\nu''+\eta_2X''+\tilde{\mathcal{E}}_2(\nu,\nu',X,X',\rho,p)=0,
        \label{eq-app:EE2}
\end{align}
which shares the same coefficients $\eta_1$ and $\eta_2$ with Eq.~\eqref{eq-app:EE1}.
We thus obtain $\tilde{\mathcal{E}}_1=\tilde{\mathcal{E}}_2$,
which can be rearranged to give the equation of the form
\begin{align}
        X'=\mathcal{F}_1(\nu,X,\rho,p)\nu'+\frac{\mathcal{F}_2(\nu,X,\rho,p)}{r}.
        \label{eq-app:dXeq}
\end{align}
We then substitute Eq.~\eqref{eq-app:dXeq} to Eq.~\eqref{eq-app:EE1} to eliminate $X'$ and $X''$.
It turns out that this procedure also removes $\nu''$ in Eq.~\eqref{eq-app:EE1}
(thanks to the degeneracy of the system),
leading to the equation of the form 
\begin{align}
        \nu'=\mathcal{F}_3(\nu,X,\rho,\rho',p).
        \label{eq-app:dnueq}
\end{align}
The explicit expressions for
$\mathcal{F}_\lambda$, $\mathcal{F}_1$, $\mathcal{F}_2$, and $\mathcal{F}_3$
are given by
\begin{widetext}
\begin{align}
    \mathcal{F}_\lambda&=
    \frac{e^{-\nu } \left[r X' \left(A_3 X+f_X\right)+2
   f\right] \left\{r X' \left[A_3 \left(3 e^{\nu }
   X-2\right)+3 e^{\nu } f_X\right]+2 f e^{\nu }
   \left(r \nu '+1\right)\right\}}{2 f \left(2 f+p
   r^2\right)},
\end{align}
$\mathcal{F}_1=2fe^{\nu}W_1/W_0$, $\mathcal{F}_2=2fe^{\nu}W_2/W_0$, and
$\mathcal{F}_3=2W_3/(r\widetilde W_0)$, with
\begin{align}
   W_0&=A_3^2 \left[4 f e^{\nu } X \left(6 e^{\nu }
   X-5\right)+p r^2 \left(15 e^{2 \nu } X^2-9 e^{\nu
   } X-2\right)-\rho  r^2 \left(3 e^{2 \nu } X^2-5
   e^{\nu } X+2\right)\right]
   \notag \\ & \quad 
   +A_3 e^{\nu } f_X
   \left\{4 f \left(6 e^{\nu } X-1\right)+r^2 \left[p
   \left(24 e^{\nu } X-5\right)+\rho  \left(5-6
   e^{\nu } X\right)\right]\right\}+3 e^{2 \nu } r^2
   f_X^2 (3 p-\rho ),
   \\ 
   W_1&=-A_3 \left[e^{\nu } X \left(8 f-\rho  r^2\right)+p
   r^2 \left(5 e^{\nu } X+1\right)+\rho 
   r^2\right]-e^{\nu } r^2 f_X (3 p-\rho ),
   \\ 
   W_2&=A_3 \left[f \left(4-8 e^{\nu } X\right)-r^2 \left(5
   e^{\nu } p X-p+\rho -e^{\nu } \rho 
   X\right)\right]-e^{\nu } r^2 f_X (3 p-\rho ),
   \\ 
   \widetilde W_0&=
   A_3^3 e^{\nu } \biggl\{
   e^{2 \nu } X \left[X f_X
   \left(256 f^2+96 f r^2 (9 p-5 \rho )+r^4 \left(395
   p^2-398 p \rho +71 \rho ^2\right)\right)-4 f
   \left(8 f+r^2 (5 p-\rho )\right)^2\right]
   \notag \\ & \quad 
   -2 e^{\nu
   } r^2 (p+\rho ) \left[2 f \left(8 f+r^2 (5 p-\rho
   )\right)+X f_X \left(r^2 (27 \rho -50 p)-36
   f\right)\right]+r^2 f_X (p+\rho ) \left[12 f+5 r^2
   (5 p-\rho )\right]\biggr\}
   \notag \\ & \quad 
   +A_3 e^{2 \nu } r^2 f_X
   (3 p-\rho ) \biggl\{e^{\nu } \left[32 f^2 X A_{3
   X}+4 f r^2 \left(X A_{3 X} (5 p-\rho )+f_{XX}
   (\rho -3 p)\right)+r^2 f_X^2 (15 p-17 \rho
   )\right]
   \notag \\ & \quad 
   +4 f r^2 A_{3 X} (p+\rho )\biggr\}
   +A_3^2
   e^{2 \nu } r^2 \biggl\{
   e^{\nu } \biggl[f_X \bigl(X
   f_X \left[24 f (9 p-7 \rho )+r^2 \left(231 p^2-284
   p \rho +61 \rho ^2\right)\right]
   \notag \\ & \quad 
   -4 f (3 p-\rho )
   \left(8 f+r^2 (5 p-\rho )\right)\bigr) 
   -4 f X
   f_{XX} (3 p-\rho ) \left(8 f+r^2 (5 p-\rho
   )\right)\biggr]
   \notag \\ & \quad 
   +2 (p+\rho ) \left[2 f r^2 f_{XX}
   (\rho -3 p)+3 f_X^2 \left(2 f+r^2 (4 p-3 \rho
   )\right)\right]\biggr\}
   \notag \\ & \quad 
   +A_3^4 \biggl\{
   4 e^{2 \nu }
   r^2 X^2 (p+\rho ) \left[47 f+r^2 (25 p-9 \rho
   )\right]+3 e^{3 \nu } X^3 \left[32 f+3 r^2 (5 p-3
   \rho )\right] \left[8 f+r^2 (5 p-\rho )\right]
   \notag \\ & \quad 
   +3
   e^{\nu } r^2 X (p+\rho ) \left[20 f+r^2 (15 p+\rho
   )\right]+6 r^4 (p+\rho )^2\biggr\}
   +4 f e^{3 \nu }
   r^4 A_{3 X} f_X^2 (\rho -3 p)^2,
   \\
   W_3&=
   -2 e^{3 \nu } f (\rho -3 p)^2 A_{3 X} f_X^2 r^4
   -e^{2\nu } (3 p-\rho ) A_3 f_X \biggl\{
   8 \left(2 e^{\nu }
   X-1\right) A_{3 X} f^2+2 r^2 \bigl[(\rho -p) A_{3
   X}
   \notag \\ & \quad 
   +e^{\nu } \left(X (5 p-\rho ) A_{3 X}+(\rho -3
   p) f_{XX}\right)\bigr] f
   -e^{\nu } r^2 f_X^2
   \left(-30 p+10 \rho +3 r \rho'\right)
   \biggr\}
   r^2
   \notag \\ & \quad 
   -e^{2 \nu } A_3^2 \biggl\{2 \left[2 (3 p-\rho )
   \left(7 r^2 \rho -6 f\right)+r \left((7 p-4 \rho )
   r^2+2 f\right) \rho'\right] f_X^2+2 f
   \left[(p-\rho ) r^2+4 f\right] (3 p-\rho )
   f_{XX}
   \notag \\ & \quad 
   -e^{\nu } \bigl[2 f X \left((5 p-\rho )
   r^2+8 f\right) (3 p-\rho ) f_{XX}+f_X \bigl(
   2 (3
   p-\rho ) \left(f \left((5 p-\rho ) r^2+8
   f\right)-4 X \left((17 p-4 \rho ) r^2+12 f\right)
   f_X\right)
   \notag \\ & \quad 
   +3 r X \left((11 p-3 \rho ) r^2+8
   f\right) f_X \rho'\bigr)
   \bigr]
   \biggr\}
   r^2-A_3^4 \biggl\{
   -2 (p+\rho ) \left[4 (p+\rho )+r
   \rho'\right] r^4
   \notag \\ & \quad 
   -e^{\nu } X \left[4 \left((15
   p-7 \rho ) r^2+24 f\right) (p+\rho )+r \left(7
   (p-\rho ) r^2+20 f\right) \rho'\right] r^2
   \notag \\ & \quad 
   -4
   e^{2 \nu } X^2 \left[2 (p-4 \rho ) (5 p-\rho )
   r^4+\left(2 r^2 (\rho -3 p)-11 f\right) \rho'
   r^3+60 f (p-\rho ) r^2+48 f^2\right]
   \notag \\ & \quad 
   +3 e^{3 \nu }
   X^3 \left((5 p-\rho ) r^2+8 f\right) \left(16
   f-r^2 \left(-20 p+4 \rho +r \rho'\right)\right)
   \biggr\}
   \notag \\ & \quad 
   -e^{\nu } A_3^3 \biggl\{f_X
   \left[r \left(r^2 (7 \rho -3 p)-4 f\right) \rho'
   -2 \left((17 p-13 \rho ) r^2+12 f\right)
   (p+\rho )\right] r^2
   \notag \\ & \quad 
   +e^{2 \nu } X \bigl[-3 X
   \left((13 p-3 \rho ) r^2+16 f\right) f_X \rho'
   r^3-2 f \left((5 p-\rho ) r^2+8 f\right)^2
   \notag \\ & \quad 
   +2 X
   \left((61 p-17 \rho ) (5 p-\rho ) r^4+24 f (19 p-5
   \rho ) r^2+64 f^2\right) f_X\bigr]
   \notag \\ & \quad 
   +2 e^{\nu }
   \bigl[X f_X \left(r (19 p-8 \rho ) \rho'-2
   \left(6 p^2-61 \rho  p+15 \rho ^2\right)\right)
   r^4+f \left((p-\rho ) \left(r^2 (5 p-\rho )-108 X
   f_X\right)+24 r X f_X \rho'\right) r^2
   \notag \\ & \quad 
   +32
   f^3+4 f^2 \left(r^2 (7 p-3 \rho )-8 X
   f_X\right)\bigr]\biggr\}.
\end{align}
\end{widetext}


One can thus reduce the number of derivatives of the TOV system in DHOST theories.
For given central values $\rho_c$ and $\nu_c$,
one can integrate Eqs.~\eqref{eq-app:dXeq},~\eqref{eq-app:dnueq},
and the hydrodynamical equation~\eqref{eq:hydro-r}, equipped with the equation of state,
to determine $\nu(r)$, $X(r)$, $\rho(r)$, and $p(r)$.
The remaining metric function $\lambda$ can then be determined from Eq.~\eqref{eq-app:soln-lambda}.
The resultant metric and $X$ are required to be matched smoothly to the external solution
at the surface $r=R$ of the fluid body, where $p(R)=0$.
For this to be possible, $\nu_c$ must be adjusted to a suitable value.
As a result, one finds a sequence of solutions parametrized by $\rho_c$.

\bibliography{refs}
\bibliographystyle{JHEP}
\end{document}